\documentclass[english,10pt, conference, letterpaper]{IEEEtran}
\usepackage[T1]{fontenc}
\usepackage[latin9]{inputenc}
\usepackage{amsthm}
\usepackage{amstext}
\usepackage{amssymb}
\usepackage{graphicx}

\makeatletter

\providecommand{\tabularnewline}{\\}

\theoremstyle{plain}
\newtheorem{thm}{\protect\theoremname}
\theoremstyle{definition}
\newtheorem{defn}[thm]{\protect\definitionname}
\theoremstyle{plain}
\newtheorem{cor}[thm]{\protect\corollaryname}
\theoremstyle{plain}
\newtheorem{prop}[thm]{\protect\propositionname}
\theoremstyle{plain}
\newtheorem{lem}[thm]{\protect\lemmaname}

\makeatother

\usepackage{babel}
\providecommand{\corollaryname}{Corollary}
\providecommand{\definitionname}{Definition}
\providecommand{\lemmaname}{Lemma}
\providecommand{\propositionname}{Proposition}
\providecommand{\theoremname}{Theorem}

\begin{document}
\global\long\def\abs#1{}

\author{\IEEEauthorblockN{Eli A. Meirom} \IEEEauthorblockA{Dept. of Electrical Engineering\\ Technion - Israel Institute of Technology\\ Haifa, Israel} \and \IEEEauthorblockN{Shie Mannor} \IEEEauthorblockA{Dept. of Electrical Engineering\\ Technion - Israel Institute of Technology\\ Haifa, Israel} \and \IEEEauthorblockN{Ariel Orda} \IEEEauthorblockA{Dept. of Electrical Engineering\\ Technion - Israel Institute of Technology\\ Haifa, Israel}}

\title{Formation Games of Reliable Networks}
\maketitle
\begin{abstract}
We establish a network formation game for the Internet's Autonomous
System (AS) interconnection topology. The game includes different
types of players, accounting for the heterogeneity of ASs in the Internet.
We incorporate reliability considerations in the player's utility
function, and analyze static properties of the game as well as its
dynamic evolution. We provide dynamic analysis of its topological
quantities, and explain the prevalence of some ``network motifs''
in the Internet graph. We assess our predictions with real-world data.
\end{abstract}

\section{Introduction}

The Internet is a primary example of a large-scale, self organized,
complex system. Understanding the processes that shape its topology
would provide tools for engineering its future.

The Internet is assembled out of multiple Autonomous Systems (ASs),
which are contracted by different economic agreements. These, in turn,
compose the routing pathways among the ASs. With some simplifications,
we can represent the resulting network as a graph, where two nodes
(ASs) are connected by a link if traffic is allowed to traverse through
them. The statistical properties of this ``Internet graph'', such
as the clustering properties, degree distribution etc., have been
thoroughly investigated \cite{Gregori2013}.  However, without proper
understanding of the mechanism that led to this structure, the statistical
analysis alone lacks the ability to either predict the future evolution
of the Internet nor to shape its evolution.

A large class of models, a primary example is the ``preferential
attachment'' model \cite{Barabasi1999}, use probabilistic rules
in order to simulate the network evolution and recover some of its
statistical properties. Yet, these models fail to account for many
other features of the network \cite{1019306}. Possibly, one of the
main reasons for that is they treat the ASs as passive elements rather
than economic, profit-maximizing entities. Therefore, an agent-based
approach is a promising alternative.

Game theory is one of the main tools of the trade in estimating the
performance of distributed algorithms \cite{Borkar2007}. It describes
the behavior of interacting rational agents and the resulting equilibria.
Game theory has been applied extensively to fundamental control tasks
in communication networks, such as flow control \cite{Altman1994},
network security \cite{Roy2010}, routing \cite{Orda1993} and wireless
network design \cite{Charilas2010}.

Recently, there has been increased activity in the field of \emph{network
formation games}. These studies aim to understand the network structure
that results from interactions between rational agents \cite{Johari2006,Jackson1996}.
Different authors emphasized different contexts, such as wireless
networks \cite{5062080} or the inter-AS topology \cite{Anshelevich2011,Alvarez2012}.
The theme of such research is to investigate the equilibria's properties,
e.g., establishing their existence and obtaining bounds on the ``price
of anarchy'' and ''price of stability''. These metrics measure from
above and below, correspondingly, the social cost deterioration at
an equilibrium compared with a (socially) optimal solution. Alternatively,
agent-based simulations are used in order to obtain statistical characteristics
of the resulting topology \cite{Lodhi2012a}.

Nonetheless, the vast majority of these studies assume that the players
are identical, whereas the Internet is a heterogeneous mixture of
various entities, such as CDNs, minor ISPs, tier-1 ASs etc. There
are only a few studies that have explicitly considered the effects
of heterogeneity on the network structure. Some examples include \cite{Alvarez2012},
which extends a previous model of formation games for directed networks
\cite{Johari2006}, and in the context of social networks, \cite{Vandenbossche2012}.
The latter describes a network formation game in which the link costs
are heterogeneous and the benefit depends only on a player's nearest
neighbors (i.e., no spillovers). 

Most of the studies on the application of game theory to networks,
with very few exceptions, e.g., \cite{Arcaute2013}, focused on static
properties of the game. This is particularly true for network formation
games. However, it is not clear that the Internet has reached an equilibrium.
Indeed, ASs continuously draw new contracts, some merge with others
while other quit business. In fact, a dynamic inspection of the inter-AS
network suggests that the system may be far from equilibrium. Therefore,
a \emph{dynamic} study of an inter-AS network formation game is needed.

In addition, previous work ignores an important requirement that Autonomous
Systems has - reliability. Indeed, failures occur, and an AS must
face such events. While some game theoretic works addressed reliability
in other contexts \cite{Bala,Haller2005}, to the best of our knowledge,
there are no works that considered the \emph{topological properties
}that emerge in a \emph{heterogeneous, dynamic,} network formation
game with \emph{reliability constraints.} 

We establish an analytically-tractable model, which explicitly accounts
for the heterogeneity of players as well as reliability requirements.
We base our model on the heralded Fabrikant model \cite{Fabrikant2003,Corbo2005},
which was recently extended to include heterogeneous players \cite{Meirom2014}.
We model the inter-AS connectivity as a network formation game with
\emph{heterogeneous} players that may share costs by monetary transfers.
We account for the inherent bilateral nature of the agreements between
players, by noting that the establishment of a link requires the agreement
of both nodes at its ends, while removing a link can be done unilaterally.
As reliability comes into play, agents may require to be connected
to other agents, or to all the other agents in the network, by at
least two disjoint paths. We investigate both the static properties
of the resulting game as well as its dynamic evolution. 

Game theoretic analysis is dominantly employed as a ``toy model''
for contemplating about real-world phenomena. It is rarely confronted
with real-world data. In this study we go a step further from traditional
formal analysis, and we do consider real inter-AS topology data analysis
to support our theoretical findings. 

The main contributions of our study are as follows:
\begin{itemize}
\item In the context of network formation games, we provide a theoretical
framework that introduces reliability constraints. We discuss both
the case of frequent failures, where the fall-back pathways are as
frequently used as the main pathways, as well as the case of rare
crashes.
\item We introduce the concept of \textquotedbl{}\emph{price of reliability}\textquotedbl{},
which is defined as the ratio of the social cost with reliability
constraints to the social cost with no such additional constraints.
Surprisingly, we show that this price can be smaller than one, namely,
that the additional reliability requirements may increase the social
utility.
\item We provide dynamical analysis of topological quantities, and explain
the prevalence of some ``\emph{network motifs}'', i.e., sub-graphs
that appear frequently in the network. Through real-world data, we
provide encouraging support to our predictions.
\end{itemize}
In the next section, we describe our model. We discuss alternative
variants that address different failure frequencies or whether utility
transfers (e.g., monetary transfers) are allowed or not; we address
both the case of allowing utility (i.e., monetary) transfers as well
as the case where this is not possible. Next, in Section 3, we provide
static analysis. Dynamic analysis is presented in Section 4. In section
5 we compare our theoretical predictions with real-world data on inter-AS
topologies. Finally, conclusions are presented in Section 6.

\section{Model}

We assume that each AS is a player. While there are many types of
players, following \cite{Meirom2014}, we aggregate them into two
types: \emph{major league} (or t\emph{ype-A}) players, such as major
ISPs, central search engines and the likes, and \emph{minor league}
(or \emph{type-B}) players, such as local ISP or small enterprises.
Each player, regardless of its type, may form contracts with other
players, and should they reach a mutual understanding, a link between
them is formed. A player's strategy is set by specifying which links
it is interested in establishing, and, if permissible, the price it
will be willing to pay for each. In order to maintain reliable routing
pathways, players may be required to sustain at least \emph{two disjoint
paths} to other players or a subset of players.

We denote the set of type-A (type B) player by $T_{A}$ ( $T_{B}$).
A link connecting node $i$ to node $j$ is denoted as either $(i,j)$
or $ij$. The total number of players is $N=|T_{A}|+|T_{B}|$, and
we assume $N\geq3$. The \emph{shortest distance} between nodes $i$
and $j$ is the minimal number of hops along a path connecting them
and is denoted by $d(i,j)$. Finally, The degree of node $i$ is denoted
by $deg(i)$.

\subsection{Basic model}

Our cost function is based on the cost structure in \cite{Fabrikant2003}
and \cite{Corbo2005}. Players are penalized for their distance from
other players. First and foremost, players require a good, fast connection
to the major players, while they may relax their connection requirements
to minor players. Bandwidth usage and delay depends heavily on the
hop distance, and connection quality is represented by this metric.
Similarly to \cite{Meirom2014}, we weight the relative importance
of a major player by a factor $A>1$ in the cost function in the corresponding
distance term. The link prices represent factors such as the link\textquoteright s
maintenance costs, bandwidth allocation costs etc. Different player
types may incur different link costs, $c_{A},c_{B}$, due to varying
financial resources or infrastructure.

All ASs must maintain access to the Internet in case of a single link
failure. This is tantamount to the requirement that all the players
must have at least two\emph{ disjoint} paths to each other node. Nevertheless,
if either link prices are high, crash frequencies are low or the content
of a minor AS is of little value, players may relax their reliability
requirements and demand the establishment of disjoint paths only to
the major players. This is represented in the cost function by a control
parameter $\tau$, which is set to one if two disjoint paths are required
to all nodes, and zero if the requirement holds for (other) nodes
of major players only. Conversely, if failures are often, then the
regular and backup paths (in the corresponding pair of disjoint paths)
are used almost as frequently. As such, they must be weighted the
same in the cost function. Therefore, the distance cost is composed
of two terms, one represents the distance along the primary path and
the other represents the distance along the backup path. The relative
weight of these two terms is set by a parameter $\delta.$ If failures
are frequent and the likelihood of using either route is the same,
we have $\delta=1$. However, if failures are rare, traffic will be
mostly carried across the shorter path. Therefore its length should
carry more weight in the cost than the length of the backup route,
hence $\delta\ll1$. This motivates the following cost function.
\begin{defn}
Two paths, $R_{(i,j)}=(i,x_{1},x_{2},...j)$ and $R'_{(i,j)}=(i,x'_{1},x'_{2},...j)$
are disjoint if they have no node in common, namely if the unordered
sets satisfy 
\begin{eqnarray*}
\left\{ x_{1},x_{2},...\right\} \cap\left\{ x'_{1},x'_{2},...\right\}  & = & \emptyset.
\end{eqnarray*}

The cost function $C(i)$ of node $i$ of type $\beta\in\{A,B\}$,
is defined as:\label{cost-definition}
\begin{eqnarray*}
C_{\beta}(i) & \triangleq & deg(i)\cdot c_{\beta}+\frac{A}{1+\delta}\sum_{j\in T_{A}}\left(d(i,j)+\delta d'(i,j)\right)\\
 &  & +\tau\cdot\frac{1}{1+\delta}\sum_{j\in T_{B}}\left(d(i,j)+\delta d'(i,j)\right)\\
 &  & +\left(1-\tau\right)\sum_{j\in T_{B}}d(i,j)
\end{eqnarray*}

where $d(i,j)$ and $d'(i,j)$ are the lengths of a pair of disjoint
paths between $i,j$ that minimizes the cost function. $d(i,j)$ denotes
the length of the shorter path. Formally, denote a pair of disjoint
paths connecting player $i$ and player $j$ as $\left(R_{(i,j)},R'_{(i,j)}\right)_{\alpha}$,
where $d_{\alpha}(i,j)$ $\left(d'_{\alpha}(i,j)\right)$ is the length
of shorter (correspondingly, longer) path. Set 
\[
\left(\hat{R}_{(i,j)},\hat{R'}_{(i,j)}\right)=\arg\min_{\left(R_{(i,j)},R'_{(i,j)}\right)_{\alpha}}C_{\beta}(i)
\]
 then $d(i,j)=\left\Vert \hat{R}_{(i,j)}\right\Vert $ and $d'(i,j)=\left\Vert \hat{R'}_{(i,j)}\right\Vert $. 

If there is not pair of disjoint path player $i$ and player $j$,
then $d'(i,j)=\mathcal{Q}$, with $\mathcal{Q\rightarrow\infty}.$If
there is not a path connecting players $i$ and $j$ we also have
$d(i,j)=\mathcal{Q}.$
\end{defn}
For convenience, we set $c\triangleq\left(c_{A}+c_{B}\right)/2$.
We assume $c_{A}\leq c_{B}.$ The social cost is the sum of individual
costs, $\mathcal{S}=\sum C_{\beta}(i)$. We denote the optimal (minimal)
social cost as $\mathcal{S}_{optimal}$\emph{, }and the social cost
at the optimal stable solution is $\tilde{S}{}_{optimal}$. The \emph{price
of stability} is the ratio between the social cost at the best stable
solution and its value at the optimal solution, namely $PoS=\tilde{S}{}_{optimal}/\mathcal{S}_{optimal}$.
Similarly, denote by $\tilde{S}{}_{pessimal}$ the highest social
cost in an equilibrium. Then, the \emph{price of anarchy} is the ratio
between the social cost at the worst stable solution and its value
at the optimal solution, namely $PoA=\tilde{S}{}_{pessimal}/\mathcal{S}_{optimal}$. 

Note that the requirement of disjoint \emph{node} paths generalizes
the requirement of disjoint \emph{link} paths and protects against
link failures within an Autonomous System. All of our results apply
to both notions of disjoint path (except a refinement of Theorem \ref{thm:symmetric dynamic};
see the discussion there). For simplicity and generality we shall
use the notion of a disjoint node paths.
\begin{defn}
We denote the change in cost of player $i$ as after the addition
(removal) of a link $(j,k)$ by $\Delta C(i,E+jk)\triangleq C\left(i,E\cup(j,k)\right)-C\left(i,E\right)$
(correspondingly, $\Delta C(i,E-jk)\triangleq C\left(i,E\right)-C\left(i,E\setminus(j,k)\right)$)
. The abbreviation $\Delta C(i,jk)$ is often used. 
\end{defn}
If $\delta=1$, then the two routing pathways are used the same. In
this case, the shortest cycle length $d(i,j)+d'(i,j)$ is the relevant
quantity that appears in the cost function. This can be found in polynomial
time by using Suurballe's algorithm \cite{Suurballe1974,Bhandari1999}.
However, if $\delta\ll1$, routing will occurs along two disjoint
paths, such that the length of the shortest between the two is shortest
(among all pairs of disjoint paths). Although the complexity of finding
this pair is NP-Hard, first finding the shortest path and then finding
the next shortest path is a heuristic that works remarkably well,
both in the real-world data analysis and on the networks obtained
in the theoretical discussion. The reason behind this is that, when
failures are rare, information is predominantly routed along the shortest
path. When players are required to establish a fall-back route, they
will establish a path that is disjoint from the \emph{current }routing
path, namely the shortest one.

The establishment of a link requires the bilateral agreement of the
two parties at its ends, while removing a link can be done unilaterally.
This is known as a \emph{pairwise-stable }equilibrium\cite{Jackson1996,Arcaute2013}. 
\begin{defn}
The players' strategies are \emph{pairwise-stable }if for all $i,j\in T_{A}\cup T_{B}$,
the following hold:

a) if $ij\in E$, then $\Delta C(i,E-ij)>0$;

b) if $ij\notin E$, then either $\Delta C(i,E+ij)>0$ or $\Delta C(j,E+ij)>0$. 

The resulting graph is referred to as a \emph{stabilizable} graph.
\end{defn}
The additional reliability requirements result in additional link
expenses, as for example, the degree of every node needs to be at
least two. The \emph{price of reliability} is the ratio between the
optimal social cost under the additional survivability constraint
to the optimal social cost when the additional constraints are removed.
\begin{defn}
The cost function,$C_{\beta}^{(bare)}(i)$, of node $i$ of type $\beta\in\{A,B\}$,
is obtained by setting $\delta=0,\tau=0$ in Definition \ref{cost-definition}
and requiring the existence of only a \emph{single} path from player
$i$ to any other players in the network. Denote the optimal social
cost without the additional survivability requirement in a pairwise
stable equilibrium as $\tilde{S}_{optimal}^{(bare)}$. The \emph{price
of reliability }(\emph{PoR)} is the ratio between the optimal value
of the social costs among the set of stable equilibria, $PoR=\tilde{S}_{optimal}/\tilde{S}_{optimal}^{(bare)}$.
\end{defn}
Surprisingly, we shall show that there exist scenarios in which reliability
requirements \emph{increase }the social utility, so that the price
of reliability can be smaller than one.

\subsection{Utility transfer}

Thus far, it was implicitly assumed that utility transfer is not feasible.
Nevertheless, often players are able to transfer utility, for example
via monetary transactions. An extended model that incorporates such
transfers is introduced by allowing for a monetary transaction in
which player $i$ pays player $j$ some amount $P_{ij}$ iff the link
$(i.j)$ is established \cite{Meirom2014}. Player $j$ sets some
minimal price $w_{ij}$ and should $P_{ij}\geq w_{ij}$ the link is
formed. 
\begin{defn}
\label{monetary cost definition}The cost function of player $i$
when monetary transfers are allowed is $\tilde{C}(i)\triangleq C(i)+\sum_{j,ij\in E}\left(P_{ij}-P_{ji}\right)$.
\end{defn}

We recall the observation in \cite{Meirom2014} that, without transfers,
a link will be established only if \emph{both} parties, $i$ and $j$,
reduce their costs, $C(i,E+ij)<0$ and $C(j,E+ij)<0$. But, when monetary
transfers are allowed, an edge will be established if (and only if)
the relaxed condition $\Delta C(i,E+ij)+\Delta C(i,E+ij)<0$ holds.
In game theoretic terms, this condition is equivalent to the requirement
that the \emph{core} of the two players game is non-empty. 

\emph{}

\emph{\begin{cor}
\textup{\emph{\label{lem:edges with monetary transfers.}When monetary
transfers are allowed, the link $(i,j)$ is established iff $\Delta C(i,E+ij)+\Delta C(j,E+ij)<0$.
The link is removed iff $\Delta C(i,E-ij)+\Delta C(j,E-ij)>0$.}}\end{cor}
}In the remainder of the paper, whenever monetary transfers are feasible,
we will state it explicitly, otherwise the basic model (without transfers)
is assumed.

\section{\label{sec:Static-analysis}Static analysis}

We shall now analyze the properties of stable equilibria, such as
the \emph{price of anarchy}, which is the ratio between the social
cost at the worst stable solution and its value at the optimal solution,
and the \emph{price of stability}, which is the ratio between the
social cost at the best stable solution and its value at the optimal
solution. We shall further discuss topological properties that emerge
from our analysis.

It was shown in \cite{Fabrikant2003} that if $c<1$ the only stable
solution is a clique and in \cite{Meirom2014} it was shown that if
$c_{A}<A$ then the major players form a clique. One may have guessed
that reliability requirements, which generally induce the creation
of additional, backup edges, would ease the formation of the clique.
The next proposition shows that this naive assumption is wrong, and
in fact, as the frequency of failure increases, it becomes increasingly
difficult to maintain the major player's clique. Consider a dense
set, in which every player may access all the other players within
two hops by a at least two disjoint paths. A direct link between two
players only reduces their mutual distance by one, and does not affect
any other distance. If this link fails often, it may be used only
partially, and it may not be worthy to pay its cost. Hence, in this
setting, counter intuitively, frequent failures end up with a sparser
network.
\begin{prop}
Assume the frequency of failures is high, namely $\delta=1.$ Then,
the type-A players form a clique if and only if $c_{A}<A/2$. Allowing
monetary transfers does not change the result. \end{prop}
\begin{IEEEproof}
We consider a major player's clique and ask under which conditions
the removal of a link is a worthy move. Consider an edge $(i,j)$
in this clique. Since only the shortest distance between players $i$
and $j$ is affected, and is increased by one, the type-A players
clique is stable if and only if $A/\left(1+\delta\right)<c_{A}$.
\end{IEEEproof}
As the major players (tier-1 AS) form a densely connected set, a clique-like
subgraph, in the rest of the paper we shall only consider the case
where $c_{A}<A/2$. We also assume, trivially, that $c_{B}>1$, as
otherwise the only stabilizable network is a clique.

 The next proposition describes a scenario in which, surprisingly,
the additional reliability constraints \emph{reduce} the social cost. 

\begin{figure}
\centering{}\includegraphics[width=0.6\columnwidth]{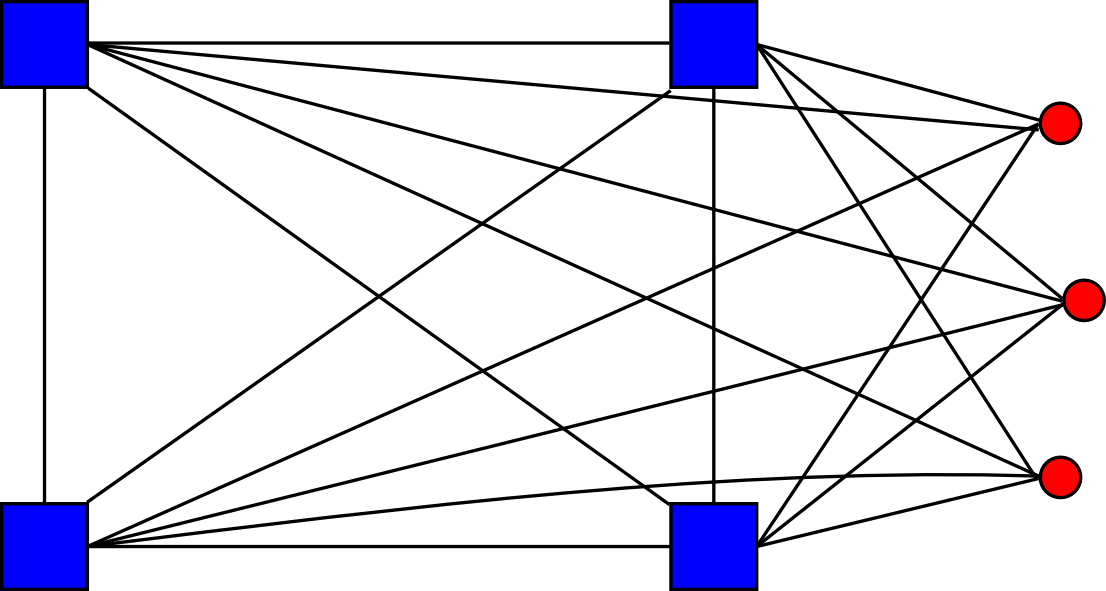}\protect\caption{\label{fig:The-optimal-solution}The optimal solution when $1<c_{A}<A/2$
and $\tau=1$, according to Prop. \ref{lem:reliable optimal 1}. The
type-B players are in red circles. The type-A clique is in blue squares.}
\end{figure}

\begin{prop}
\label{lem:reliable optimal 1} Assume $1<c_{A}<A/2$ and symmetric
reliability requirements, namely $\tau=1$. Then, the optimal network
is composed of a type-A clique, where all the type $B$ nodes are
connected to all members of the type-A clique, as depicted in Fig.
\ref{fig:The-optimal-solution}. This network is not stabilizable,
and $PoS>1$. Nevertheless, for $|T_{B}|\gg|T_{A}|\gg1$, we have
$PoS\rightarrow1$. In addition, the Price of Reliability is smaller
than one. 
\end{prop}
The main idea behind this result is that in the optimal, yet unstable
solution, every minor player establishes a link with all the major
players (Fig. \ref{fig:The-optimal-solution}). This configuration
is unstable as it is over-saturated with links, and the optimal stable
solution  is obtained by diluting this network so that every minor
player will connect to just two major players. If the reliability
requirements are further removed, then  additional dilution occurs,
increasing the social cost. In other words, the stable configuration
is under-saturated with edges, and the additional survivability requirements
facilitate the formation of additional links. 
\begin{IEEEproof}
� � We shall now prove the first part of this theorem, namely that
the network described in Fig. \ref{fig:The-optimal-solution}, is
optimal in terms of social cost. In this network, every type-B player
is connected to every type-A player. We denote this network by $G.$

First, consider a network in which the type-A players are not in a
clique. Then, there exists two players $i,j\in T_{A}$ such that $d(i,j)\geq2$.
By establishing a link $(i,j)$ the change in social cost is
\[
\Delta\mathcal{S}\leq-2\left(\frac{A}{1+\delta}-c_{A}\right)<0.
\]
Therefore a network which includes this link has a lower social cost.
Hence, in the optimal network, all the type-A players are in clique.
A similar calculation for a missing link between player $i\in T_{A}$
and $j\in T_{B}$ shows that
\[
\Delta\mathcal{S}\leq-\left(\frac{A+1}{1+\delta}-c_{A}-c_{B}\right)<0
\]

and establishing this link reduces the social cost as well. In conclusion,
in an optimal network a type-A player is connected to all the other
players.

Consider node $i\in T_{B}.$ Its distance from every other node $j\in T_{B}$
is $d(i,j)=2$ and $d'(i,j)=2$. The minimal distances between two
minor players $i,j\in T_{B}$ are $d(i,j)=1$ and $d'(i,j)=2$ whereas
in $G$ we have $d(i,j)=d'(i,j)=2$. Assume that a network with a
lower social cost exists, and denote it by $G'$. In $G'$ there must
exists a link between two type-B players, $i,j\in T_{B}$. However,
removing this link reduces the social cost, as

\[
S(E+ij)-S(E)=2\left(\frac{A}{1+\delta}-c\right)>0
\]
Therefore, $G'$ is not optimal, in contradiction to our assumption.
This proves that $G$ is optimal.

Next, we are going to show $G$ is unstable. Consider $i\in T_{B}$
and $x\in T_{A}$. By removing $(i,x)$ we cost change of player $x$
is 
\[
\Delta C(x,E-ix)=-2\left(\frac{1}{1+\delta}-c\right)>0
\]

And therefore it is beneficial for player $x$ to remove this link.
Hence, this network in not stabilizable.

Finally, we are going to show that the $PoR<1$. The outline is as
follows. First, we are going to show that the network depicted in
Fig. $\ref{fig:optimal and stable}$ is stable. The optimal stable
network (without reliability requirements) in this parameter regime
was explicitly derived in \cite{Meirom2014}. We are going to show
that the cost in the latter network is higher than the former, which
bound the $PoR$ from above by a value smaller than one.

Consider a network in which all the minor players are connected to
two major players, $x,y\in T_{A}$. We are going to show that this
network is stable. Clearly, as shown before, the type-A clique is
stable. In addition, for any player $i\in T_{B}$ neither $x,y$ nor
$i$ has the incentive to remove either $(x,i)$ or $(y,i)$ as it
would violate the reliability requirement and would lead to unbounded
cost. Hence, this network is stable. Denote this network by $\tilde{G}.$
In \cite{Meirom2014} it was shown that the optimal stable network
(without reliability requirements) in this configuration is the network
in which all type-A players form a clique, and all the type-B players
are connected to a single type-A player. We denote the latter network
by $G''.$We have
\[
\mathcal{S}(\tilde{G}.)-S(G'')=|T_{B}|\left(\frac{A+1}{1+\delta}-c_{A}-c_{B}\right)<0.
\]

But, 
\[
PoR=\frac{\tilde{S}_{optimal}}{\tilde{S}_{optimal}^{(bare)}}\leq\frac{S(\tilde{G}.)}{\mathcal{S}(G'')}=1+\frac{\mathcal{S}(\tilde{G}.)-S(G'')}{\mathcal{S}(G'')}<1.
\]

This concludes the proof.
\end{IEEEproof}
In conclusion, if the failure frequency is high, the survivability
requirements will induce dilution of the clique of major players.
However, the opposite effect occurs along the graph cut-set between
the set of minor players and the set of major players, where the additional
constraints lead to an increased number of links connecting major
and minor players.

So far we have assumed that the reliability requirements are symmetric.
As explained in the Introduction, in some cases it is reasonable to
assume that players will require a backup route only to the major
players, i.e., non-symmetric reliability constraints. We shall now
show that in this case, the social cost may deteriorate considerably.
Hence, from a system designer point of view, it is much more important
to incentivize a configuration where the reliability requirements
are symmetric, than to reduce failure frequency globally. This result
stems from the following lemma.
\begin{lem}
\label{lem:finite distances}If $\tau\neq0$ then for every two nodes
i,j, $d(i,j)\leq d'(i,j)\leq2+2c_{B}$. If $\tau=0$ there exists
stable equiliberia such that there are no two disjoint paths between
some nodes $i,j$.\end{lem}
\begin{IEEEproof}
First, we are going to show that $d'(i,j)$ is finite (step 1). Then,
in the second step, we shall bound $d'(i,j)$ from above.

Note, that if there aren't two disjoint paths between players $i$
and $j$, it is beneficial for both players to add the link $(i,j)$
and form a path.

Step 1: If $i$ and $j$ are disconnected, then it is worthy for both
of them to establish the link $(i,j)$. Therefore, there exists a
path from $i$ to $j$ and $d(i,j)$ is finite. Assume $d(i,j)\neq1$.
If there aren't two disjoint paths between nodes $i$ and $j,$ it
is worthy for both nodes to establish a direct link. Hence if $d(i,j)\neq1$
the distance $d'(i,j)$ is finite. We now discuss the case $d(i,j)=1$
and show there must exists an additional disjoint path, so $d'(i,j)$
is finite. We prove by negation.

Take node $x\neq i,j$. As before, both $d(i,x)$ and $d(x,j)$ are
finite. Consider the following two cases:

A) $d(i,x)=1$, $d(x,j)=1$. Then, the trajectory $(i,x,j)$ is disjoint
from the path $(i,j)$, and it is therefore worthy for both player
$i$ and player $j$ to establish the link $(i,j).$ Hence $d'(i,j)$
is finite.

B) $d(i,x)=1$, $d(x,j)\neq1$ or $d(i,x)\neq1$, $d(x,j)=1$. Without
loss of generality, we assume the first case. According to the previous
discussion, there are at least two disjoint paths from $x$ to $j.$
Since they are disjoint, only one of them may contain the edge $(i,j)$.
Therefore, there exist a path $(i,x,...j)$ that is disjoint from
the edge $(i,j)$ and therefore $d'(i,j)$ is finite.

C) $d(i,x)\neq1$, $d(x,j)\neq1$ . Applying the same reasoning as
in B), there exists a path from $i$ to $x$ that is disjoint from
the edge $(i,j)$. Likewise, there exists a path from $j$ to $x$
that is disjoint from the edge $(i,j).$ Therefore there exist a path
from $i$ to $j$ that is disjoint from $(i,j).$

Step 2: Step 1 has shown that every two nodes are connected by a cycle.
Let us assume the longest shortest cycle connecting two nodes is of
length $l>2+2c_{B}>3$, namely there exists a cycle $(x_{0},x_{1},x_{2},...,x_{\left\lfloor l/2\right\rfloor },...x_{l},x_{0})$.

By establishing the link $(x_{0},x_{\left\lfloor l/2\right\rfloor })$
the cost of player $x_{0}$ due to distances from other players is
reduced by at least $\left(l^{2}-1+mod(l+1,2)\right)/4$, hence the
link will be established as
\[
\Delta C\left(x_{0},E+(x_{0},x_{\left\lfloor l/2\right\rfloor }\right)\leq\left(l^{2}-1+mod(l+1,2)\right)-c_{B}\leq0
\]

and similarly $\Delta C\left(x_{0},E+(x_{0},x_{\left\lfloor l/2\right\rfloor }\right)\leq0$
. Since $d(i.j)\leq d'(i.j)\leq l$ the proof is complete.
\end{IEEEproof}
If there exist players with no two disjoint paths in an equilibrium,
then the social cost becomes unbounded. This immediately results in
an \emph{unbounded} Price of Anarchy, as indicated by the following
theorem.
\begin{thm}
\label{thm:price of anarchy}Consider a network such that $T_{B}\gg T_{A}\gg1$.
The Price of Anarchy is as follows:

A) In a setting with asymmetric reliability requirements $(\tau=0)$:
unbounded.

B) In a setting with symmetric reliability requirements $(\tau\neq0)$:
bounded by $o(c)$. \end{thm}
\begin{IEEEproof}
A) We shall prove this by showing that there exists a stable equilibrium
with unbounded social cost. Consider a network in which the major
players (type-A players) form a clique, while every minor player is
connected only to a single type-A player $j\in T_{A}$. We shall show
that this network is stabilizable. As discusses before, the type-A
clique is stable. Consider $i\in T_{B}.$ By forming the link $(i,x)$
the change of cost of player $x\in T_{A}$, $x\neq j$ is
\[
\Delta C(x,ix)=c_{A}-1>0.
\]

Hence, this link will be formed. No additional links between minor
players will be established, since such a link only reduces the distance
between the participating parties by one, $c_{B}-1>0$ and does not
provide an additional disjoint path to the type-A clique. Therefore,
the social cost is $\mathcal{S=\omega}(\mathcal{Q})\rightarrow\infty$. 

B) The total cost due to the inter-connectivity of the type-A clique
is identical for all link stable equilibria and is $|T_{A}|\left(|T_{A}|-1\right)\left(c+\left(1+\delta/2\right)A\right)$.
This cost is composed of $|T_{A}-1|$ links per node, and the distance
cost to every other major nodes, 
\[
d(i,j)+d'(i,j)=1+\delta/2.
\]
Next, we evaluate the cost due to the type-B nodes inter-distances.
According to Lemma \ref{lem:finite distances}, both $d(i,j)\leq4c_{B}$
and $d'(i,j)\leq4c_{B}$, so the cost due to the inter-distances between
a type-B player and every other player is bounded from above by $4c_{B}\left(|T_{A}|+|T_{B}|\right)$.
When summed up over all minor players, this contributes a term $4c_{B}|T_{B}|\left(|T_{A}|+|T_{B}|\right)$
to the social cost. Likewise, the cost of links that at least one
of their ends is a type B player is at most $c_{B}|T_{B}|\left(|T_{A}|+|T_{B}|\right)$. 

Therefore, the maximal cost in all link stable equilibria is bounded
from above by 
\[
|T_{A}|\left(|T_{A}|-1\right)\left(c+\left(1+\delta/2\right)A\right)+5c_{B}|T_{B}|\left(|T_{A}|+|T_{B}|\right).
\]

The optimal network configuration is described in Proposition \ref{lem:reliable optimal 1}.
It is straightforward to evaluate the social cost in this configuration.
The distance cost due to inter-distances of the type-B players is
\[
2|T_{B}|\left(|T_{B}-1\right),
\]

while the cost of all links which connect type-B players to type-A
players is
\[
|T_{B}||T_{A}|\left(c_{B}+c_{A}\right).
\]

Finally, the social cost due to the type-A clique remains the same,
so finally we obtain that the minimal social cost is

\begin{eqnarray*}
\mathcal{S}_{optimal} & = & |T_{A}|\left(|T_{A}|-1\right)\left(c+\left(1+\delta/2\right)A\right)\\
 &  & +2|T_{B}|\left(|T_{B}-1\right)+2|T_{B}||T_{A}|\left(c_{B}+c_{A}\right)
\end{eqnarray*}

By taking the limits $|T_{A}|\rightarrow\infty,|T_{B}|/|T_{A}|\rightarrow\infty$
we obtain that $PoA\leq5c_{B}/2$.
\end{IEEEproof}
A stable equilibrium with infinite social cost can be easily achieved
by considering a network where all minor players are connected to
a single, designated, major player. There exists a single path of
at most two hops between every minor player to every major player.
However, as the stability requirements are asymmetric, the major players
have no incentive to establish additional routes to any minor player,
and the reliability requirements of the minor players remain unsatisfied.

\subsection{Monetary transfers }

\label{sec:Monetary-transfers}

The previous discussion assumed that a player cannot compensate other
players for an increase in their costs. Yet, contracts between ASs
often do involve monetary transactions. Accordingly, in this subsection
we shall highlight the additional insights that are obtained when
utility transfers are permissible.

Our first result indicates that, in this setting, in contrast to the
previous setting, there always exists a fallback route between every
two players, regardless of the symmetric or asymmetric nature of the
additional survivability constraints. If monetary transfers are feasible,
players may compensate other players for the cost of additional links
such that all the additional constraints are satisfied. Hence, symmetry
is less important than in the previous scenario. Furthermore, this
result suggests that every player is connected to every other player
by a cycle. The following proposition shows that the maximal cycle
length decays with the number of major players. As the number of ASs
increases in time, this predicts that this length should decrease
in time. We shall verify this prediction in Section \ref{sec:Data-Analysis}.
\begin{prop}
\label{prop:max distance}Assume $1<c<A/2$ . Then, every two players
are connected by a cycle, and the maximal cycle length connecting
a major player to a minor player is bounded by 
\[
\max\left\{ 2\left(\left\lfloor \sqrt{\left(A|T_{A}|\right)^{2}+5c}-A|T_{A}|\right\rfloor +1\right),4\right\} .
\]
\end{prop}
\begin{IEEEproof}
\textit{\emph{Lemma \ref{lem:finite distances} showed that the maximal
distance between players is bounded. We are going to tighten this
result in the regime where monetary transfers are feasible. Denote
the maximal distance between type-A player and a type\_B player by
$k_{A}$.}}

\textit{\emph{First, we are going to show that the maximal cycle length
connecting a major player and a minor player is $2k_{A}+1.$ This
follows from a simple geometric argument. Consider two players, $i\in T_{A}$
and $j\in T_{B}$. Assume $k_{A}>2$. If the cycle length is $2k_{A}+2$
or greater, there exists a type-B node that its distance $k_{A}+1,$
in contradiction to the assumption that the maximal distance between
a major player and and a minor player is $k_{A}$. Denote maximal
distance between two minor players by $k_{B}.$ A similar argument
shows that maximal cycle length between two minor players is $2k_{B}+1$.}}

\textit{\emph{Next, are going to show that the maximal distance connecting
a major player and a minor player is $k_{A}\leq l\leq\max\left\{ 2\left\lfloor \sqrt{\left(A|T_{A}|\right)^{2}+5c}-A|T_{A}|\right\rfloor ,2\right\} $
. We prove by negation. Assume that the distance between player $j\in T_{A}$
and $i\in T_{B}$ is $l$. Denote the nodes on the path as $(x_{0}=i,x_{1},x_{2}....,x_{l}=j)$.
Then, by establishing a link between them, the distance between $j$
and $\{x_{0},x_{1}....x_{\left\lfloor l/2\right\rfloor }\}$ (similarly,
and distance between player $i$ and players $\{x_{\left\lceil l/2\right\rceil }...x_{l-1}\}$)
is reduced. In addition, player $i$ reduces its distance to every
node of the type-A clique by $l$. Lemma 25 of \cite{Meirom2014}
shows that the total reduction in distance is $\left(l^{2}-1+mod(l+1,2)\right)/4$.
Then, by establishing the link $(i,j)$ we have
\[
\Delta C(i,E+ij)+\Delta C(j,E+ij)\leq c_{A}+c_{B}-2\left(l^{2}-1\right)-l|T_{A}|
\]
}}

\textit{\emph{and as $l\geq2\left\lfloor \sqrt{\left(A|T_{A}|\right)^{2}+5c}-A|T_{A}|\right\rfloor $
this expression is negative. Therefore, the link will be established,
and the maximal shortest distance between a major player and a minor
player is smaller than $l$. This concludes the proof.}}
\end{IEEEproof}
Our second result is based on the first one, and shows that the price
of anarchy is bounded. In fact, as the network grows, Proposition
\ref{prop:max distance} also indicates that its diameter \emph{shrinks}.
Therefore, in the large network limit, the price of anarchy is bounded
by a constant.
\begin{prop}
The price of anarchy is bounded by $o(c)$. Furthermore, if $|T_{B}|\gg1,|T_{A}|\gg1$
then the price of anarchy is upper bounded by 2. \end{prop}
\begin{IEEEproof}
In Proposition \ref{prop:max distance}'s proof we showed that the
maximal distance between a major player a minor player is bounded
by 
\[
l\leq\max\left\{ 2\left\lfloor \sqrt{\left(A|T_{A}|\right)^{2}+5c}-A|T_{A}|\right\rfloor ,2\right\} 
\]

It immediately follows that in the large network limit, as $l\leq2$,
every player is directly connected to a player that, in turn is connected
to the type-A players. The latter can be either a major player, as
part of the clique, or a minor player that is connected to every player
in the major player's clique. Note that a link between any two players
$y_{i}$ and $y_{j}$ reduces the social cost, since it at least lowers
the costs of $y_{i}$ of $y_{j}$ and can only reduce the costs of
other players. Therefore, we can bound the social cost by a configuration
in which every type-B player has the minimal number of links, namely
two, is at distance two from every major player.

Therefore, the worst social cost in bounded by
\begin{eqnarray*}
\mathcal{S} & \leq & |T_{A}|\left(|T_{A}|-1\right)\left(c+\left(1+\delta/2\right)A\right)\\
 &  & +2c|T_{B}|+2|T_{B}|\left(|T_{B}-1\right)
\end{eqnarray*}

Comparing this bound with $\mathcal{S}_{optimal}$, as derived in
Proposition \ref{lem:reliable optimal 1}, in the limit $|T_{A}|\rightarrow\infty,|T_{B}|/|T_{A}|\rightarrow\infty$
completes the proof.

\end{IEEEproof}

\section{\label{sec:Dynamic-Analysis}Dynamic Analysis}

The Internet undergoes continuous transformations, such as the emergence
of new ASs, or formation of new traffic contracts. In fact, it may
very well be out of equilibrium. Therefore, a static analysis of the
equilibrium points must be accompanied by dynamic analysis. Accordingly,
our main focus in this section is to identify prevalent network motifs
\cite{Milo2002a}, i.e., small sub-graphs that emerge during the natural
evolution of the network. In Section \ref{sec:Data-Analysis} we shall
show that these motifs are indeed ubiquitous in the real AS topology,
and the frequency of their occurrences is few folds\emph{ }more than
expected in a random network.

While there are many possible equiliberia, we shall show that convergence
occurs only to \emph{just a few}. We shall also show that the convergence
time is short, namely \emph{linear }in the number of players.

We start the discussion by setting up the dynamic framework, as first
formulated in \cite{Meirom2014}.

\subsection{Setup \& Definitions}

We split the game into \emph{turns}, where at each turn only a single
player is allowed to remove or initiate the formation of links. At
each point in time, or turn, the players that already joined the game
form a subset $N'\subset T_{A}\cup T_{B}$. We shall implicitly assume
that the cost function is calculated with respect to the set $N'$
of players that are already present in the network. Each turn is divided
into \emph{moves, }at each of which a player either forms or removes
a single link. A player's turn is over when it has no incentive to
perform additional moves. Note that disconnections of several links
can be done unilaterally and hence iteratively. 
\begin{defn}
Dynamic Rule \#1: In player $i$'s turn it may choose to move $m\in\mathcal{N}$
times. In each move, it may remove a link $(i,j)\in E$ or, if player
$j$ agrees, it may establish the link $(i,j)$. Player $j$ would
agree to establish $(i,j)$ iff $C(j;E+(i,j))-C(j;E)<0$.
\end{defn}
According to this definition, during player's $i$ turn, all the other
players will act in a greedy, rather than strategic, manner. For example,
although it may be that player $j$ prefers that a link $(i,j')$
would be established for some $j'\neq j$, if we adopt Dynamic Rule
\#1 it will accept the establishment of the less favorable link $(i,j).$
In other words, the active player has the advantage of initiation
and the other players react to its offers. There are numerous scenarios
in which players cannot fully forecast other players' moves and offers,
e.g., when information is asymmetric or when only partial information
is available \cite{5173479}. In these settings, it is likely that
a greedy strategy will become the modus operandi of many players.
This is a prevalent strategy also when the system evolves rapidly
and it is difficult to assess the current network state and dynamics.

In a dynamic network formation game, a key question is: Can a player
temporarily disconnects itself from the graph, only to reconnect after
getting to a better bargaining position? Or must a player stay connected?
If the timescale in which the costs are evaluated is comparable to
the timescale in which the dynamics occur, then, clearly, a player
will not disconnect from the network voluntarily. However, if the
latter is much shorter, it may, for a very brief time, disconnect
itself from the graph in order to perform some strategic move. The
following rules address the two alternative limits. 

\begin{defn}
Dynamic Rule \#2a: Let the set of links at the current move $m$ be
denoted as $E_{m}$. A link $(i,j)$ will be added if $i$ asks to
form this link and $C(j;E_{m}+ij)<C(j;E_{m})$. In addition, any link
$(i,j)$ can be removed in move $m.$

Dynamic Rule \#2b: In addition to Dynamic Rule \#2a, player $i$ would
only remove a link $(i,j)$ if $C(i;E_{m}-ij)>C(i;E_{m})$ and would
establish a link if both $C(j;E_{m}+ij)<C(j;E_{m})$ and $C(i;E_{m}+ij)<C(i;E_{m})$\end{defn}

According to Dynamic Rule \#2a, a player is allowed to perform a strategic
plan in which the first few steps will increase its cost, as long
as when the plan is completed its cost will be reduced.  On the other
hand, if the game follows Dynamic Rule \#2b, then a player's cost
must be reduced \emph{at each move}, hence such multi-move plan is
not possible.

\subsection{\label{sub:dynamical Results}Basic Model - Results}

\begin{figure}
\centering{}\includegraphics[width=1\columnwidth]{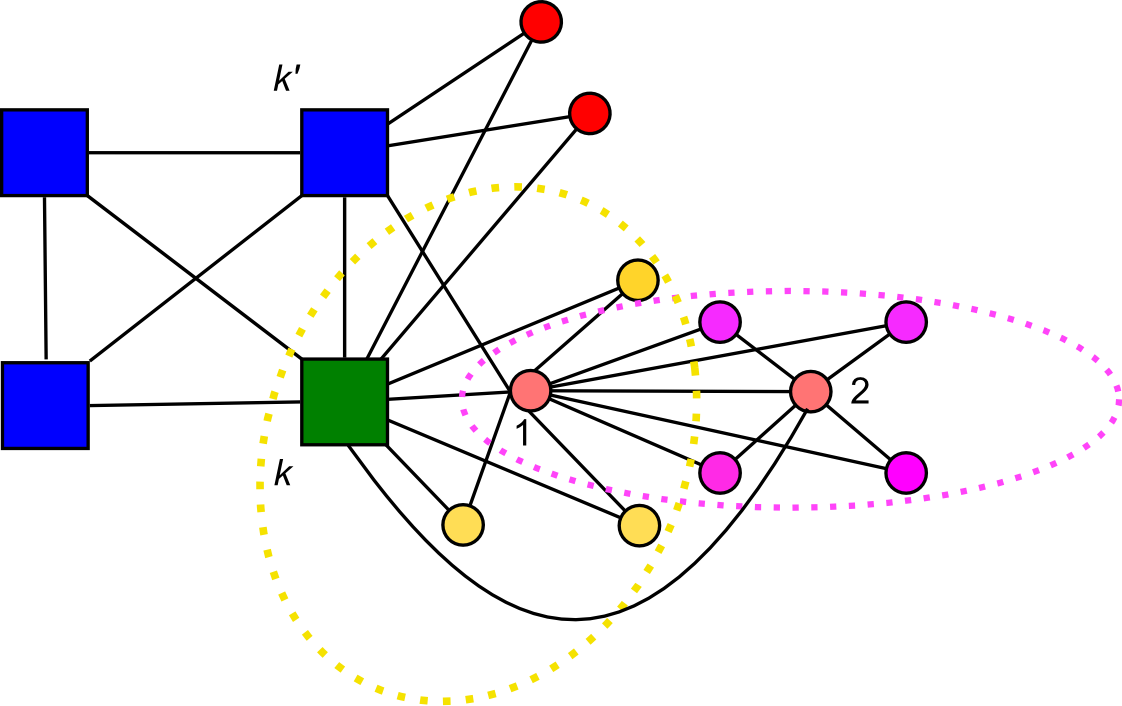}\protect\caption{\label{fig:generalized star 2}A network configuration which includes
a ``double-star'' structure of minor players. Every node in the
primary star (encircled in yellow) is linked to a major player, node
$k$ (in green). A direct link connects the two star centers, denoted
by 1 and 2 (in pink). The members in the secondary star (in purple)
are connected to both star centers. In addition, there secondary star
center is also connected to the major player $k$. There might be
additional minor players outside the stars (in red). Needless to say,
the type-A clique (square boxes) is also present. }
\end{figure}

Our first result in this section shows that, during the natural evolution
of the network, a ``double star'' sub-graph, or network motif, often
emerges. In the ``double star'' motif, as depicted in Fig. \ref{fig:generalized star 2},
there exists a primary and a secondary star. All the minor players
are connected to the primary star's center. Part of the players are
also connected to the other star's center, forming the secondary star.
Consider a region where it is immensely difficult to establish a link
to a major player, either due to geographical distance, link prices
or perhaps additional physical links are simply not accessible. Nevertheless,
in order to maintain a reliable connection, there must be at least
two links that connect this region to the Internet backbone via some
major players. In order to provide a stable, fault tolerant service,
every player in this region will form links with the players hosting
the endpoints of these links, forming the double star sub-graph. Assume
now that link prices reduce over time, or that the importance of a
fast connection to the Internet core increases in time. In this case,
players may decide to establish direct links with the major players,
and remove either one of both links connecting them to the star centers.
Note though, that players will be reluctant to disconnect from the
star center if the number of nodes in the star is large.

\begin{figure}
\centering{}\includegraphics[width=0.6\columnwidth]{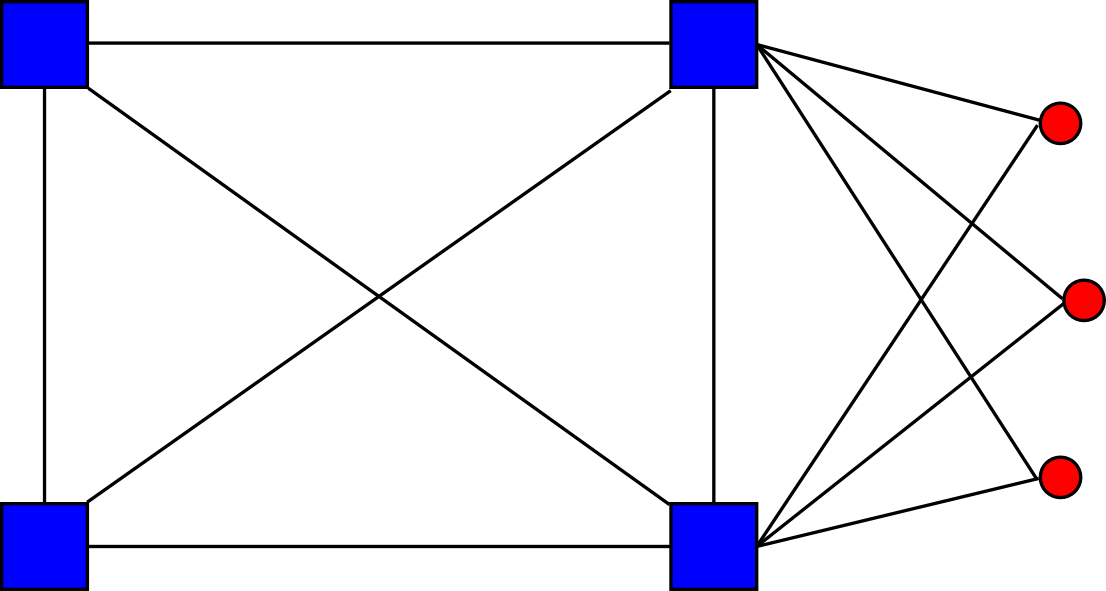}\protect\caption{\label{fig:optimal and stable}The optimal stable network, as described
in Theorem \ref{thm:symmetric dynamic}. }
\end{figure}

Moreover, the next theorem also shows that, eventually, and fairly
quickly, the system will converge to either the optimal stable state,
or to a state in which the social cost is a low multiple of the optimal
social cost.
\begin{thm}
\label{thm:symmetric dynamic}Assume symmetric reliability requirements,
i.e., $\tau=1$. If the players follow Dynamic Rules \#1 and \#2a,
then, in any playing order:

A) The system converges to either the optimal stable state, depicted
in Fig. \ref{fig:optimal and stable}, or to the network depicted
in Fig. \ref{fig:generalized star 2}.

B) In the large network limit, namely, when $|T_{B}|\gg|T_{A}|\gg1$,
the social costs ratio satisfy $\mathcal{S}/\mathcal{S}_{optimal}<3/2+\epsilon$,
with $\epsilon\rightarrow0$.

C) If players play in a uniformly random order, the probability that
the system has not converged by turn $t$ decays exponentially with
$t$. Otherwise, if every player plays at least once in O(N) turns,
convergence occurs after O(N) steps.\end{thm}
\begin{IEEEproof}
We claim that the network, at any time, has the following structure
(Fig. \ref{fig:generalized star 2}): A type-A players' clique, and
\emph{at most} two type-B star centers $1,2\in T_{B}$. The larger
ball's center is labelled as $1$, and the smaller is $2$. We denote
the set of type-B players which are members of the star centered about
node 1 (2) as $S_{1}$ and (correspondingly, $S_{2}$). We have $\left|S_{1}\right|\geq\left|S_{2}\right|$
In addition, some type-B players might be linked to two nodes $k,k'\in T_{A}$,
the set of these type-B players is denoted by $L$. The set of type-A
players that have a direct link with the star center $1$ (star center
$2$) is denoted by $D_{1}$ (correspondingly $D_{2}$). Players $\{1,2,k\}$
form a clique. Assuming the second star exists, it is connected to
either players $1$ or $k'$ (or both). Finally, there may be additional
links between players is $L$ and $D_{2}$. An example of this network
structure is presented in Fig. \ref{fig:generalized star 2}. 

\begin{table}
\centering{}%
\begin{tabular}{|c|c|c|c|}
\hline 
$d(i,j)$ & $i\in n_{1}$ & $i\in n_{2}$ & $i\in n_{3}$\tabularnewline
\hline 
\hline 
$j=1$ & 1 & 1 & 2\tabularnewline
\hline 
$j\in S_{1}$ & 2 & 2 & 2\tabularnewline
\hline 
$j=2$ & 2 & 1 & 2\tabularnewline
\hline 
$j\in S_{2}$ & 2 & 2 & 3\tabularnewline
\hline 
$j\in L$ & 2 & 2 & 2\tabularnewline
\hline 
$j=k$  & 1 & 2 & 1\tabularnewline
\hline 
$j=k$' & 2 & 2 & 1\tabularnewline
\hline 
$j\in D_{2}$, $j\neq k,k'$ & 2 & 2 & 2\tabularnewline
\hline 
$j\in T_{A}$, $j\notin D_{2}$ & 2 & 3 & 2\tabularnewline
\hline 
\end{tabular}\protect\caption{\label{tab:d(i,k)}The shortest distance $d(i,j)$ between two nodes,
as discussed in Prop. \ref{thm:symmetric dynamic}. Note that if $k'\in D_{2}$,
than column 9 applies instead of column 8 for $i\in S_{2}$.}
\end{table}

\begin{table}
\begin{centering}
\begin{tabular}{|c|c|c|c|}
\hline 
$d'(i,j)$ & $i\in n_{1}$ & $i\in n_{2}$ & $i\in n_{3}$\tabularnewline
\hline 
\hline 
$j=1$ & 2 & 2 & 2\tabularnewline
\hline 
$j\in S_{1}$ & 2 & 3 & 2\tabularnewline
\hline 
$j=2$ & 2 & 2 & 3\tabularnewline
\hline 
$j\in S_{2}$ & 3 & 2 & 3\tabularnewline
\hline 
$j\in L$ & 3 & 3 & 2\tabularnewline
\hline 
$j=k$  & 2 & 2 & 2\tabularnewline
\hline 
$j=k$' & 2 & 3 & 2\tabularnewline
\hline 
$j\in D_{2}$, $j\neq k,k'$ & 2 & 2 & 2\tabularnewline
\hline 
$j\in D_{1}$, $j\notin D_{2}$ & 2 & 3 & 2\tabularnewline
\hline 
$j\notin D_{1},$$j\in T_{A}$ & 3 & 3 & 2\tabularnewline
\hline 
\end{tabular}
\par\end{centering}

\centering{}\protect\caption{\label{tab:d'(i,k)}The second shortest distance $d'(i,j)$ between
two nodes, as discussed in Prop. \ref{thm:symmetric dynamic}. Note
that if $k'\in D_{2}$, than column 9 applies instead of column 8
for $i\in S_{2}$.}
\end{table}

We prove by induction. After the first three players played the induction
base case is true. We first assume that the link $(1,2)$ and consider
the case $(1,2)\notin E$ later. Denote the active player by $r$.
Consider the following cases:

1. $r\in T_{A}$: As $c<A/2$, $r$ will form (or maintain) links
with every type-A player. If additional links to minor players are
to be formed, it is clearly better for player $r$ to establish links
first with either player $1$ or player $2$ than to any player in
$i\in S_{1}$ or $i\in S_{2}$. We first consider the case where $r\notin D_{1}\cup D_{2}$
and split into two cases:

1A) If no two disjoint paths exists between player $r$ to players
$1$ or $2$, namely, if $|T_{A}|\geq2$ and either $|D_{1}|<\min\{|T_{A}|,2\}$
or $|D_{2}|\leq\min\left\{ |T_{A}|,2\right\} $, then it is beneficial
for both $r$ and the star centers to link such that reliability requirements
will be satisfied. 

1B) If $D_{2}\neq\emptyset$ then by establishing the edge $(r,1)$
we have
\begin{eqnarray*}
\Delta C(r,E+1r) & = & c-\frac{1+|S_{1}|}{1+\delta}
\end{eqnarray*}

while by establishing $(r,2)$ the change of cost is:

\begin{eqnarray}
\Delta C(r,E+2r) & = & c-\frac{1+|S_{2}|}{1+\delta}.\label{eq:2r edge}
\end{eqnarray}

As $|S_{2}|\leq|S_{1}|$, player $r$ will prefer to establish first
a link with player 1 rather than with player 2. The link will be formed
if $\Delta C(1,E+1r)\leq0$, which is true if 
\begin{eqnarray}
|D_{1}| & \leq & 1\nonumber \\
 & \mathcal{\text{or}}\label{eq:a-b link condition}\\
c_{B} & < & A/\left(1+\delta\right).\nonumber 
\end{eqnarray}
If condition \ref{eq:2r edge} holds, then player $r$ will attempt
to form a link with the star center $2$, and succeed if condition
\ref{eq:a-b link condition} holds as well. Establishing a link to
any other type-B node is clearly an inferior option, as if a player
had decided to establish links with either node $1$ or node $2,$
an additional link to one of their leafs will only reduce the distance
to it by 1 and hence is not a worthy course of action. Therefore,
$D_{2}\subseteq D_{1}$.

2. $r\in T_{B}$ and $r\neq1,2$. A player may disconnect itself and
then choose its two optimal links. Clearly, among the type-A clique,
player $r's$ best candidates are players $k,k'$, while among the
type-B players, the cost reduction is maximal by linking to nodes
$1,2$. In addition, a link to node $k$ is at least as preferred
a link to $k'$, while a link to node 1 is always preferred over a
link to node $2$ as long as $n_{1}\geq n_{2}$. Therefore, we only
need to consider three possible moves by player $r$ - linking to
$k,k'$ (the red players in Fig. \ref{fig:generalized star 2}), linking
to $k,1$ or linking to $1,2$. 

Using tables \ref{tab:d(i,k)} and \ref{tab:d'(i,k)} we can compare
the cost of connecting to nodes $1,2$ versus the costs of connecting
to nodes $1$ and $k$. A simple calculation shows that if the expression
\[
A-1+(|D_{1}|-|D_{2}|)A+\delta\frac{|S_{1}|-|S_{2}|+A+A(\abs{T_{A}}-|D_{2}|)}{1+\delta}
\]

is positive, than linking to players $1$ and $k$ is preferred over
linking to player $1$ and $2$. This expression is positive, as $|D_{1}|\geq|D_{2}$
and $|S_{2}|\leq|S_{1}|$.

In a similar fashion, we can compare between the choices of linking
to nodes $1$ and $k$ or forming links with nodes $k$ and $k'.$
If the expression 
\[
1+|S_{2}|-A+\delta\frac{1-|L|-A(\abs{T_{A}}-|D_{1}|)}{1+\delta}
\]

is positive, then establishing links to $1$ and $k$ is preferred,
otherwise the alternative will be chosen. Two special cases are of
interest: If $|S_{2}|<A-1$ and $\delta=0$, then every player will
prefer to link to $k$ and $k'$. That is, we'll get the optimal configuration.
If $\delta=1$, $T_{A}=D_{1}$, i.e, node 1 is a member of the clique,
$|L|=0$ and $|S_{2}|\gg1$, then a minor player will connect to players
$1$ and $k$. Note that if there are no type-A players present when
$r$ plays, then it must connect to the star centers $1$ and $2$. 

In addition, after $r$ formed two links, no player $x\in T_{A}\cup\{1,2\}\cup D_{1}\cup L$
will agree to form the link $(x,r)$, as the induced change of cost
is
\[
\Delta C(x,E+xr)\geq c_{A}-1>0.
\]
If $c_{B}\leq2$ and $\delta\rightarrow0$ then player $r\in L$ and
player $x\in D_{2}$ will establish the link $(x,r)$. By symmetry,
this also happens when $r\in D_{2}$ and $x\in L$.

3. If $r=1,$ then it must maintain links to all nodes in $S_{2}\cup S_{1}\cup\{2\}$
(if there are any) in order to satisfy the reliability criteria. Additional
links to type-A players will be formed according to the discussion
in case 1, while no other links to type-B player will be formed according
to the discussion in case 2. 

4. If $r=2$, then it must maintain links to all nodes in $S_{2}$
(if there are any) in order to satisfy the reliability criteria. A
similar calculation to the one in case $2$ shows that player $2$'s
best course of action is to remain connected to player 

In order to complete the induction proof all that is left is to address
the case $(1,2)\notin E$. In this case, player $2$ must be connected
to at least one additional player $i\notin S_{2}$ in order to have
two disjoint path to every player in the network. Clearly, its optimal
choice is player $k'$. In order to maintain reliable path to every
$i\in S_{2}\cup\{2\}$ player $k'$ must agree to form this link.
Player $2$ will disconnect $(1,2)$ if either $\Delta C(2,E-12)\leq0$
or $\Delta C(2,E-12+2k')\leq0$ . However, if either of this conditions
hold, Tables \ref{tab:d(i,k)} and \ref{tab:d'(i,k)} shows that for
every $i\in T_{B}/\{2\}$, it would prefer to link to player $k'$
rather than player $2$. In particular, this is also true for every
node $i\in S_{2}$. Therefore, as soon as every member in $S_{2}$
played once, the star will be empty and no additional star will raise
again. Therefore, there will be at most one star left in the network.

This completes the proof on the momentarily structure of the network.
Note that if $|S_{1}|=|S_{2}|=\emptyset$ we obtain the optimal stable
solution. 

B) Since every minor player has either two or three links, the contribution
to the social cost due to minor players' links is $o(|T_{B}|),$ while
the contribution to the social costs due to the inter-distances between
minor players is $o\left(T_{B}^{2}\right)$. In the limit $|T_{A}|\rightarrow\infty,|T_{B}|/|T_{A}|\rightarrow\infty$,
the dominant term in the cost function is the term proportional to
$|T_{B}|^{2}$. Both $d(i,j)$ and $d'(i,j)$ for $i,j\in T_{B}$
are bounded by $3$, and comparing this results with the optimal social
cost (Proposition \ref{lem:reliable optimal 1}) we have
\[
\frac{\mathcal{S}}{\mathcal{S}_{optimal}}=\frac{3T_{B}^{2}+o(T_{B}^{2})}{2T_{B}^{2}+o(T_{B}^{2})}\leq\frac{3}{2}+\epsilon
\]

with $\epsilon\rightarrow0$ in the limit $|T_{A}|\rightarrow\infty,|T_{B}|/|T_{A}|\rightarrow\infty$. 

C) The discussion of case 2 shows that in a large network, it is suboptimal
for a minor player to be a member of $S_{2}$. Given the the opportunity,
it will prefer to become a member in either $S_{1}$ or $L$. Therefore,
after every player has played at least twice, $S_{2}=\emptyset$,
and after every player has played four times the system will reach
equilibrium. Lemma 13 in \cite{Meirom2014} then shows that the probability
that the system has not converged by turn $t$ decays exponentially
with $t$. 
\end{IEEEproof}

In section \ref{sec:Static-analysis} we emphasized the importance
of symmetry in the reliability requirements for reducing the Price
of Anarchy. The next theorem affirms this assertion, and shows that
if the constraints are asymmetric, the system converges to a state
with an unbounded social cost on a large set of possible dynamics
and initial conditions.
\begin{thm}
\label{thm:dynamic asymetric}Assume asymmetric reliability requirements,
namely $\tau=0$. If the players follow Dynamic Rules \#1 and either
Dynamic Rule \#2a or \#2b, then the system converges to a state with
an unbounded social cost.\end{thm}
\begin{IEEEproof}
The idea behind the proof is to show that the reliability requirements
of at least one minor player remain unsatisfied, the therefore the
social costs, as a sum over individual costs, is unbounded.

First assume that $c_{A}>2$. We shall now show that at any given
turn, the network is composed of a type-A (possibly empty) clique,
a set of type-B players $S$ linked to player $x$, acting as a star
center, and an additional (possibly empty) set of type-B players $L$
connected to the type-A player $k$. Player $x$ is also connected
to player $k$. Some major player may establish links with the star
center $x.$ The set of these major players is denoted by $D$. Note
that the cost of every player in either $S$ or $L$ is $\omega\left(Q\right)\rightarrow\infty$,
since every path from each of these players to the type-A clique crosses
player $k$. 

We prove by induction. At turn $t\leq2$, this is certainly true.
Denote the active player at time $t$ as $r.$ Consider the following
cases:

1. $r\in T_{A}$: Since $1<c_{A}<A,$ all links to the other type-A
nodes will be established or maintained, if $r$ is already connected
to the network. Clearly, the optimal link in $r$'s concern is the
link with the star center $x$. Therefore, player $r$ will attempt
to establish the link $(r,x)$ if
\begin{equation}
\Delta C(r,E+rx)=c_{A}-|S|-1\label{eq: term1}
\end{equation}
is negative. If $c_{B}<A/\left(1+\delta\right)$ or that $|D|\leq1$
then $x$ will accept this link. Regardless, if player $r$ has formed
the link it will not establish a link with $i\in S$, as it only reduces
its distance to player $i$ by a single hop, and 
\begin{eqnarray}
\Delta C(r,E+ir) & = & c_{A}-1\leq0.\label{eq:no single hop}
\end{eqnarray}

If Eq. \ref{eq: term1} is positive, then, player $r$ has no incentive
to establish a link with any $i\in S$, as $|S|\geq1$ and the by
establishing $(i,r)$ the sole change is the reduction of $d(i,r)$
from $3$ to $1$, 
\[
\Delta C(r,E+ir)=c_{A}-2\geq c_{A}-|S|-1=\Delta C(r,E+rx)\geq0.
\]

Eq. \ref{eq:no single hop} also shows that if $r\neq k$, it will
not form a link with $i\in L$, while if $r=k$ it may not remove
the link $(k,i)$ as otherwise $i$ is disconnected.

2. $r\in T_{B},\, r\neq x$ : First, assume that $r$ is a newly arrived
player, hence it is disconnected. Obviously, in its concern, a link
to the star's center, player $x$, is preferred over a link to any
other type-B player. Similarly, a link to a player $k$ is preferred
over a link to any other type-A player (or if $L=\emptyset$ and $D=T_{A}$,
equivalent to a link to any other type-A player). Therefore, $r$
first link choice would be either $(r,k)$ or $(r,x)$. In other words,
$r\in L$ or $r\in S$. If $r\in L$, than no $i\in T_{A}$ will agree
to establish a link with $r$, as it only reduces its distance from
$r$ by one hop, and doesn't alter $d(i,x)$ for any other $x$. Similarly,
if $r\in S$ than 
\begin{eqnarray*}
\Delta C(i,E+ri) & = & c_{A}-2>0
\end{eqnarray*}

and the link would not be formed.  Likewise, no link between $i\in L$
and $j\in S$ may be formed, as 
\[
\Delta C(i,E+ij)=c_{B}-2>c_{A}-2>0
\]

3. $r=x$, the star's center: $r$ may not remove any edge connected
to a type-B player and render the graph disconnected. On the other
hand, the previous discussion shows it will not establish additional
links to nodes in $L$.

Note that the reliability requirements of players in $L\cup S$ are
invalidated. Therefore, the cost of every player $j\in L\cup S$ is
at least $\mathcal{Q}$, and we have $Q\rightarrow\infty$. Hence,
at any given turn, as soon as either $|L|>1$ and $|S|\geq1$, the
social cost is unbounded.

If $c_{A}<2$ a link between player $i\in T_{A}$ and player $r\in T_{B}$
will be formed. However, as soon as player $i$ becomes the active
player, it is beneficial for it to remove the link $(i,r)$ and establish
a link to the star center $(i,x)$ instead. Therefore, after player
$i$'s turn, the cost of player $r$ is again $\omega(Q)$. In a similar
fashion, if $c_{B}<2$ than player $i\in S$ and player $j\in L$
may establish the link $(i,j)$. This does not affect any other player
and while it may reduce their costs, it does not provide them with
two disjoint paths to the type-A clique, as they both must traverse
player $k$ in order to access it. Therefore, their costs is still
$\omega(Q)$.
\end{IEEEproof}
Although, at first sight, it might seem that this result is due to
the greedy and myopic choice of edges, it is possible to show that
there are cases in which this result holds even when players may pick
their links strategically rather than in a greedy manner. For example,
if $c_{A}>2$ and the number of minor players is odd, than a similar
analysis shows that a player in $S\cup\{x\}$ will establish a link
to a single player in $L$. However, as the number of minor players
is odd, there will always be an unmatched player in either $L$ or
$S$ and its cost will be $\omega(Q)$.

\subsection{Monetary Transfers}

Recall that under the presence of monetary transfers, players $i$
and $j$ will agree to establish an edge if $\Delta C(j,ij)+\Delta C(i,ij)<0$.
Nevertheless, it may be that during the active player turn there are
a few links that satisfy this condition, and the player\emph{ }must
prioritize them. Each player's decision is myopic, and is based solely
on the current state of the network. Hence, the order of establishing
links is potentially important. Needless to say, a player's preference
order will depend on the link prices. While there are several alternatives,
we adopt the following preference order \cite{Meirom2014}: 

Denote the active player as player $i$. Each link $(i,j)$ carries
different utility in player $i's$ respect. It is reasonable to assume
that a link with a lower \textquotedbl{}connection value\textquotedbl{}
will be priced lower, so that the link with the least connection utility
will be marked with the lowest price. In fact, one may assume its
price will be as much as the implied cost of the other party of this
link. We will denote this price as $P^{*}$. Every other player $x$
will use this value and demand an additional payment from player $i$
for the link $(i,x)$, as it is more beneficial for player $i$. Formally:

\begin{defn}
``Strategic'' Pricing mechanism: Set $j^{*}$ as the node that maximizes
$\Delta C(i,E+ij*)$. Set $P^{*}=\max\{-\Delta C(j*,E,ij*),0\}$.
Finally, set $\alpha_{ij}=\Delta C(i,E+ij)-\left(\Delta C(i,E+ij^{*})+P^{*}\right).$
The price that player $j$ requires in order to establish \emph{$(i,j)$
}is\emph{ }$P_{ij}=\max\{0,\alpha_{ij},-\Delta C(j,E+ij)\}$.
\end{defn}
Under this pricing mechanism, there could be many links that carry
the same utility. Some of these links have a better connection value,
but they come at a higher price. Since all the links carry the same
utility, we need to decide on some preference mechanism for player
$i$. The simplest one is the ``cheap'' choice, in which, if there
are a few equivalent links, the player will choose the cheapest one.
This can be a reasonable choice, as new players cannot spend too much
resources, and therefore they will choose the ``cheapest'' option
that belongs to the set of links with maximal utility.

\begin{defn}
Preference order: Player $i$ will establish links with player $j$
if player $j$ minimizes $\Delta\tilde{C}(i,ij)=\Delta C(i,ij)+P_{ij}$
and $\Delta\tilde{C}(i,ij)<0$. If there are several players that
minimize $\Delta\tilde{C}(i,ij)$, then player $i$ will establish
a link with a player that minimizes $P_{ij}$. If there are several
players that satisfy the previous condition, then one out of them
is chosen randomly.

We are now at a position to identify an additional network motif,
namely, the ``entangled cycles''. This network motif is composed
of a line (i.e., interconnected sequence) of minor players' nodes,
with some cross-links between the nodes along this line, breaking
the hierarchy (Fig. \ref{fig:The-entangled-loops}). The ``entangled
cycle'' of length three is the ``feedback loop'' motif, which was
previously found to exist in a higher frequency than expected in the
Internet graph \cite{Milo2002a}. 

\begin{figure}
\centering{}\includegraphics[width=1\columnwidth]{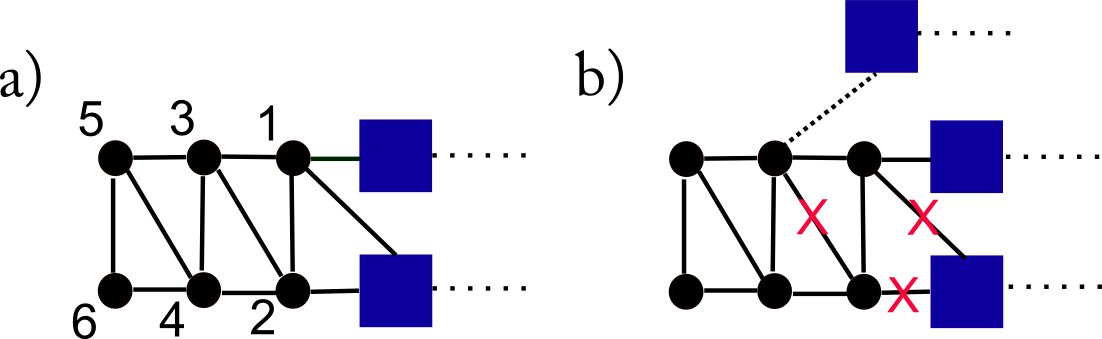}\protect\caption{\label{fig:The-entangled-loops}The ``entangled cycles'' motif.
a) Six minor player are connected in an ``entangled cycles'' subgraph.
The first two nodes have direct connection to some major player, and
access the rest of the network by the major player's additional links,
represented by the dotted line. b) If at some point, a link between
a player in this subgraph and some external player is formed (in this
example, a major player), some links may be removed without violating
the reliability requirement and without increasing the distance cost
appreciably. The removed links marked by a red X.}
\end{figure}

When a new minor player arrives, it will choose the two cheapest links
and will connect to the corresponding players. Clearly, its costs
due to the distance from the rest of the network will be the highest.
As such, when the next player arrives, it will offer the lowest link
price. The new arrival will link to it and to one of its provider.
The process will repeat, until, at some point, an existing player
will decide that this growing branch is too far from it, and will
connect to one of the nodes along this ``entangled cycles''. At
this point, this subgraph will be over-saturated with links as players
may utilize this link to access the Internet core. Hence, some links
will be removed (see, for example Fig. \ref{fig:The-entangled-loops}(b)).
The set of the links that will be removed depends heavily on the playing
order and the temporary network structure. Nevertheless, some cross-links
may remain in order to satisfy reliability constrains. This explains
the following result.\end{defn}
\begin{thm}
\label{thm:entangled cycles motif}Assume the number of major players
is at least 10. Denote the distance cost of player $i\in T_{\beta},$
$\beta\in\{A,B\}$, as $D(i)$, namely
\[
D(i)=C_{\beta}(i)-c_{\beta}deg(i)\cdot x_{i}
\]
Assume that a subset $W$ of minor player first join the game and
play consecutively and that the two players with the maximal distance
cost are adjacent. Then:

A) These players will form an ``entangled cycles'' structure of
length $l$, as depicted in Fig. \ref{fig:The-entangled-loops}, and
\[
l\leq2\sqrt{\left(A|T_{A}|\right)^{2}+5A}-2A|T_{A}|.
\]

B) The \textquotedblleft entangled cycles\textquotedblright{} structure
is semi-stable, in the following sense: If, at some later turn, there
exist players $j\notin W$, $i\in W$ such that the link $(i,j)$
is formed (Fig. \ref{fig:The-entangled-loops}(b)), then some links
in the ``entangled cycles'' structure may be removed in subsequent
turns.\end{thm}
\begin{IEEEproof}
A) Assume that at time $t$ a subset of $W=\{x_{1},x_{2}...x_{m}\}$
minor players first join the game. Denote the set of players that
are currently connected to the network by $N'$. We denote by $n_{A}$
($n_{B})$ the number of type A players (correspondingly, type-B player)
at that moment. Let us denote the players with the highest distance
cost term as $x_{0}$ and $x_{1}$, namely,
\begin{eqnarray*}
x_{0} & = & \arg\max_{i\in N'}D(i)\\
x_{-1} & = & \arg\max_{i\in N'\backslash\left\{ x_{-1}\right\} }D(i)
\end{eqnarray*}

where for simplicity we assumed that $x_{-1},x_{-2}$ are minor player
(type-B players). According to the ``strategic'' pricing mechanism
a new player will establish links with these players, as they will
offer the cheapest links. Note that by connecting to these players,
its survivability requirements are satisfied, as each of these players
maintain two disjoint path to every major player (if the reliability
requirements are asymmetric) or to all other players (in case of symmetric
reliability requirements).

We are now going to prove by induction that player $x_{j}$ will first
connect to players $x_{j-1}$ and $x_{j-2}$. We shall prove by this
by proving that the distance costs of $x_{j-1}$ and $x_{j-2}$ are
maximal.

First, note that for every player $x_{j'}=1...j-1$, we have $D(x_{j'})\geq D(x_{j'-1})$
and $D(x_{j'})>D(x_{j'-2})$, since the path from $j$' to every player
in $N'$ pass through $x_{j'-1}$ or $x_{j'-2}$, and $N'\gg l>|W|$.
Therefore, in order to show that players $x_{j-1}$ and $x_{j-2}$
have the highest distance cost, it is sufficient to show that $D(x_{j-1})>D(y)$
for every $y\in N'$. For every player $i\in N$' we have 
\begin{eqnarray*}
d(x_{j-1},i) & \geq & d(x_{0},i)+\left\lfloor j/2\right\rfloor \\
d'(x_{j-1},i) & \geq & d'(x_{0},i)+\left\lfloor j/2\right\rfloor 
\end{eqnarray*}
 since the path that connects player $x_{j-1}$ to the any player
$i\in N'$ crosses either player $x_{0}$ or $x_{-1}$, which are
adjacent. Therefore, the distance cost of player $x_{j-1}$ due to
its distance from every player $i\in N'$, denoted by $\tilde{D}(x_{j-1})$
is at least grater than the corresponding distance cost of player
$x_{0}$, $\tilde{D}(x_{0})$ by at least $A\left\lfloor j/2\right\rfloor n_{A}+\left\lfloor j/2\right\rfloor n_{B}$.
Note that $\tilde{D}(x_{0})\geq\tilde{D}(y)$ for every $y\in N'$
according to the definition of $x_{0}$. However, player $x_{j'-1}$
may be closer to players $\{x_{1},x_{2}...x_{j-2}\}$ than player
$y$. Denote the maximal distance between any two players by $r.$
We have,
\[
D(x_{j-1})-D(y)\geq A\left\lfloor j/2\right\rfloor n_{A}+\left\lfloor j/2\right\rfloor n_{B}-r\cdot j
\]

where $r$ is the maximal distance between two players. In proposition
\ref{prop:max distance} it was shown that 
\begin{eqnarray*}
r & \leq & 2\left\lfloor \sqrt{\left(A|T_{A}|\right)^{2}+5c}-A|T_{A}|\right\rfloor |\\
 & \leq & 2\sqrt{\left(A|T_{A}|\right)^{2}+5A}-2A|T_{A}|
\end{eqnarray*}

Hence, 
\[
D(x_{j-1})-D(y)\geq A\left\lfloor j/2\right\rfloor n_{A}+\left\lfloor j/2\right\rfloor n_{B}-r\cdot j
\]

As $|T_{A}|>10$ this expression is positive. Hence, links will be
established as stated. This shows that at the time player $x_{j}$
joins the game, the distance cost of player $x_{j-1}$ is maximal.
A similar calculation shows that at this turn, the distance cost $D(x_{j-2})$
is maximal in the set $\left\{ D(i)|i\in N'\cup\{x_{1},..x_{j-2}\}\right\} $. 

We have thus far shown that the first two links of a new player $x_{j}$
are to players $x_{j-1}$ and $x_{j-2}$. Nevertheless, as the ``entangled
cycles'' motif grows, the incentive of other players to connect to
players in it increases. At some point, player $j$ may be able to
form links with players not in $W$, or an active player $r\notin W$
may decide to connect to some player in $W$ (or vice versa for $r\in W$).
In either of these cases, players in the ``entangled cycles'' motif
may have three disjoint paths to other players, and may therefore
remove a link, should they deemed to do so. In this case, the ``entangled
cycles'' motif will be diluted.
\end{IEEEproof}
This theorem shows that reliability is a major factor in breaking
up tree hierarchy in the Internet. In addition, it also hints that
the hierarchical structure does not break frequently in the top levels
of the Internet, but rather mostly in the intermediate and lower tiers.
Note that according to Theorem \ref{thm:entangled cycles motif}(B),
in large networks the length of the ``entangled motifs'' is short.
Therefore, we do not expect to see excessively long structures, but
rather small ones, having just a few ASs.

\section{\label{sec:Data-Analysis}Data Analysis}

In this section we compare our theoretical predictions with the real-world
inter-AS topology graph \cite{Gregori2011}. We classified ASs to
major players and minor players according the popular CAIDA ranking
\cite{CAIDA}. For the sake of comparison with previous work \cite{Meirom2014},
we classified the major players as the top 100 ASs according to this
ranking. 

\begin{figure}
\centering{}\includegraphics[width=1\columnwidth]{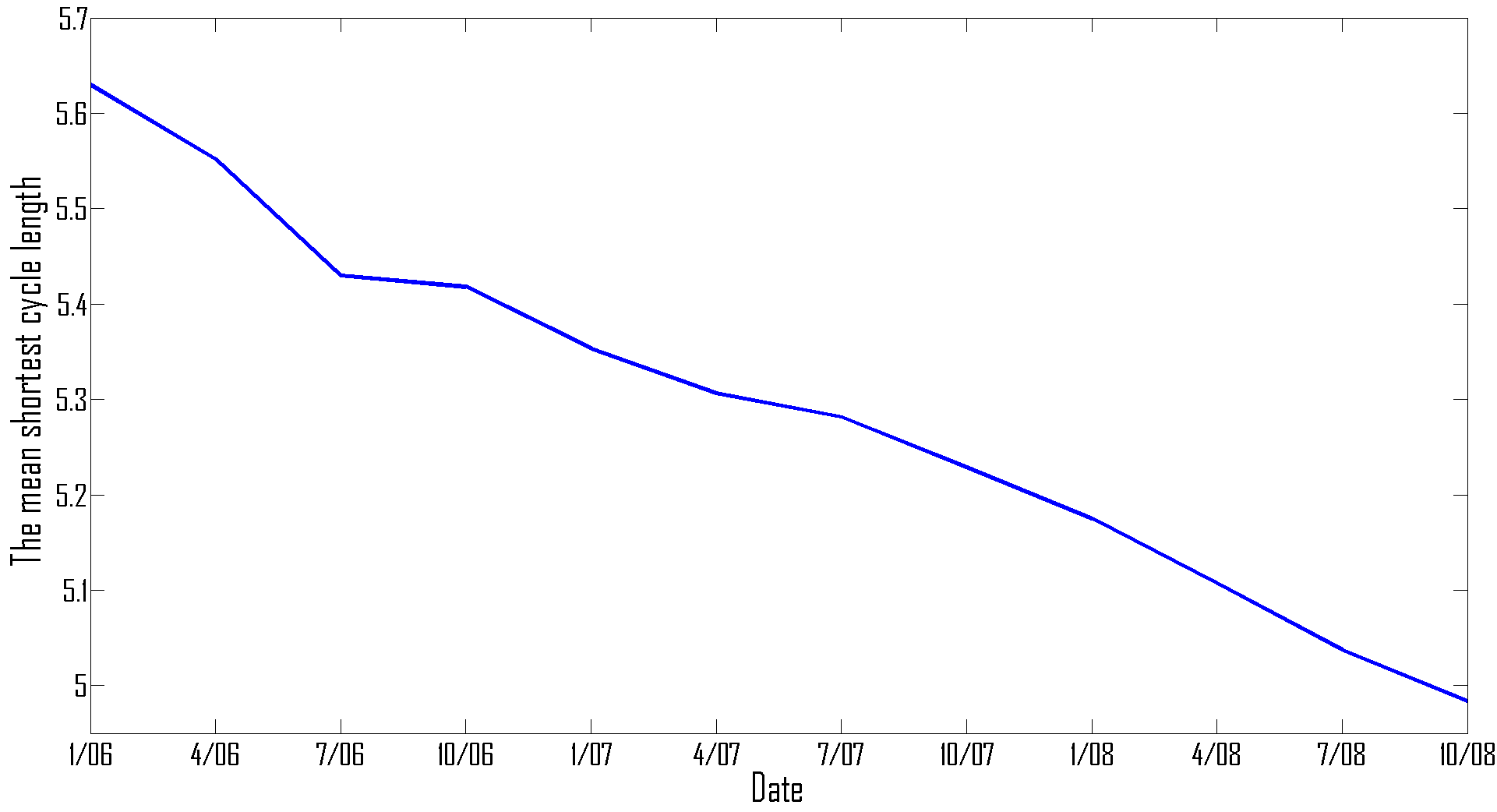}\protect\caption{\label{fig:The-mean cycle-length}The mean length of the shortest
cycle connecting a major player and a minor player as function of
time, from January 2006 to October 2008. The length decrease as time
passes and the network grows, in agreement with our model. }
\end{figure}

In Section \ref{sec:Monetary-transfers}, we showed that, by allowing
monetary transfers, the maximal cycle length connecting a major player
to a minor player depends inversely on the number of major players.
As the number of ASs increases in time, it is reasonable that the
number of major player grows as well. Hence, we expect that the length
of the shortest cycle connecting a major player to a minor player
will decrease in time. Fig. \ref{fig:The-mean cycle-length} shows
the mean cycle length connecting one of the secondary leading 2000
ASs, ranked 101-2100 in CAIDA ranking, and one of the top 100 nodes.
The steady decline of the cycle's length in time is predicted by our
model.

\begin{figure}
\centering{}\includegraphics[width=1\columnwidth]{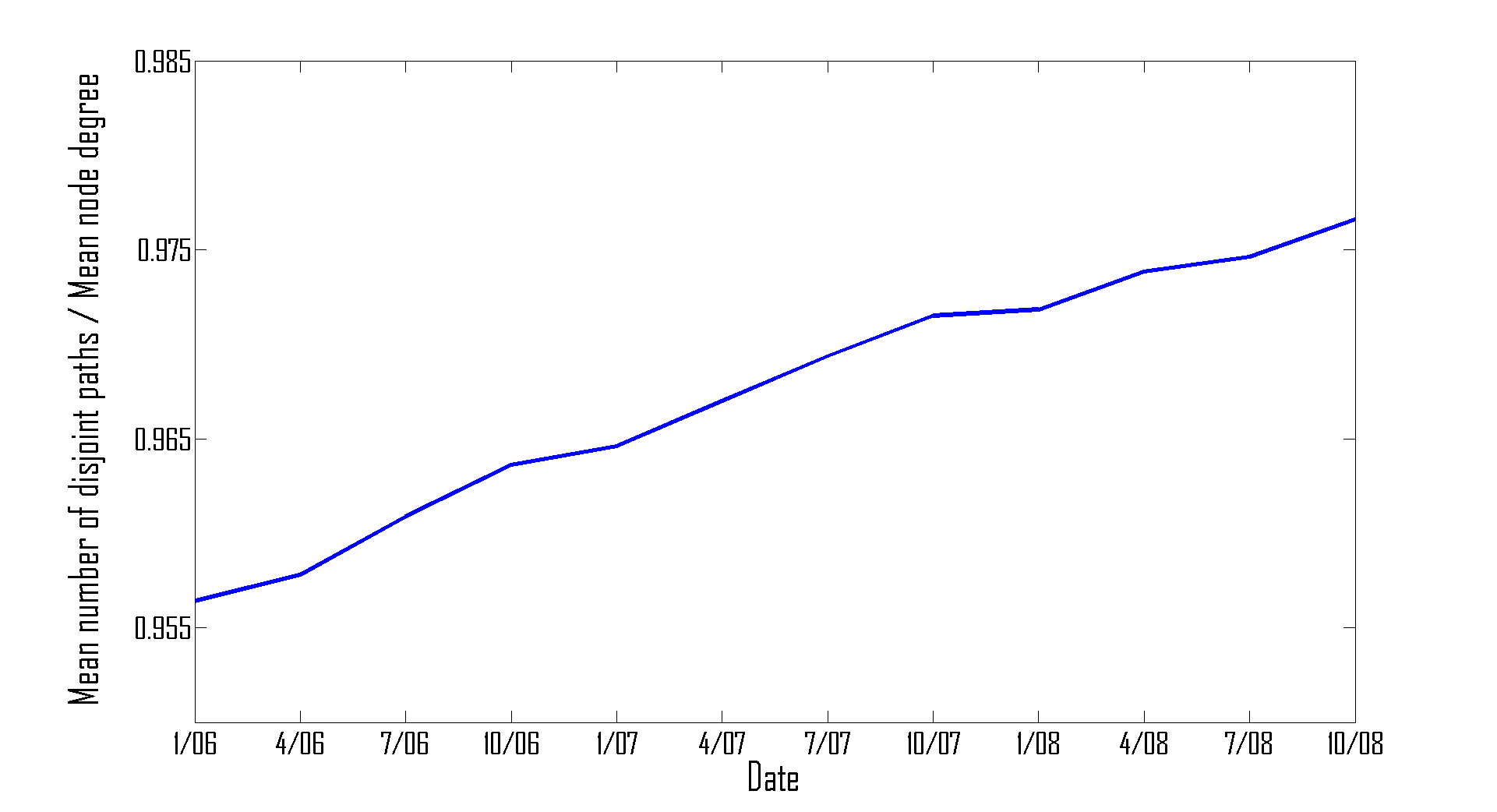}\protect\caption{\label{fig:disjoint path ratio}The ratio of the mean number of disjoint
paths connecting a minor player to the core to the mean degree of
a minor player. This ratio is above $0.95$, and increases in time,
showing that additional links are likely to part of disjoint paths
to the core. This ratio is presented as a function of time from January
2006 to October 2008.}
\end{figure}

Our analysis showed that in most of the generated topologies, the
minor players are organized in small subgraphs that have direct connection
to the Internet core, namely the major players clique, or the tier-1
subgraph, in agreement with \cite{Meirom2014}. In order to maintain
a reliable connection, in each subgraph there must be at least two
links that connect minor players to the core. Indeed, we have found
out that the ratio between the mean number of disjoint paths from
a minor player to the core and the mean degree of minor players is
more than $0.95$, and it increases in time (Fig. \ref{fig:disjoint path ratio}).
That is, almost every outgoing link of a minor player is used to provide
it with an additional, disjoint path to the core. In other words,
a player is more likely to establish an additional link, hence increase
its degree, if it supplies it with a new path to the core that does
not intersect its current paths.

\begin{figure}
\centering{}\includegraphics[width=1\columnwidth]{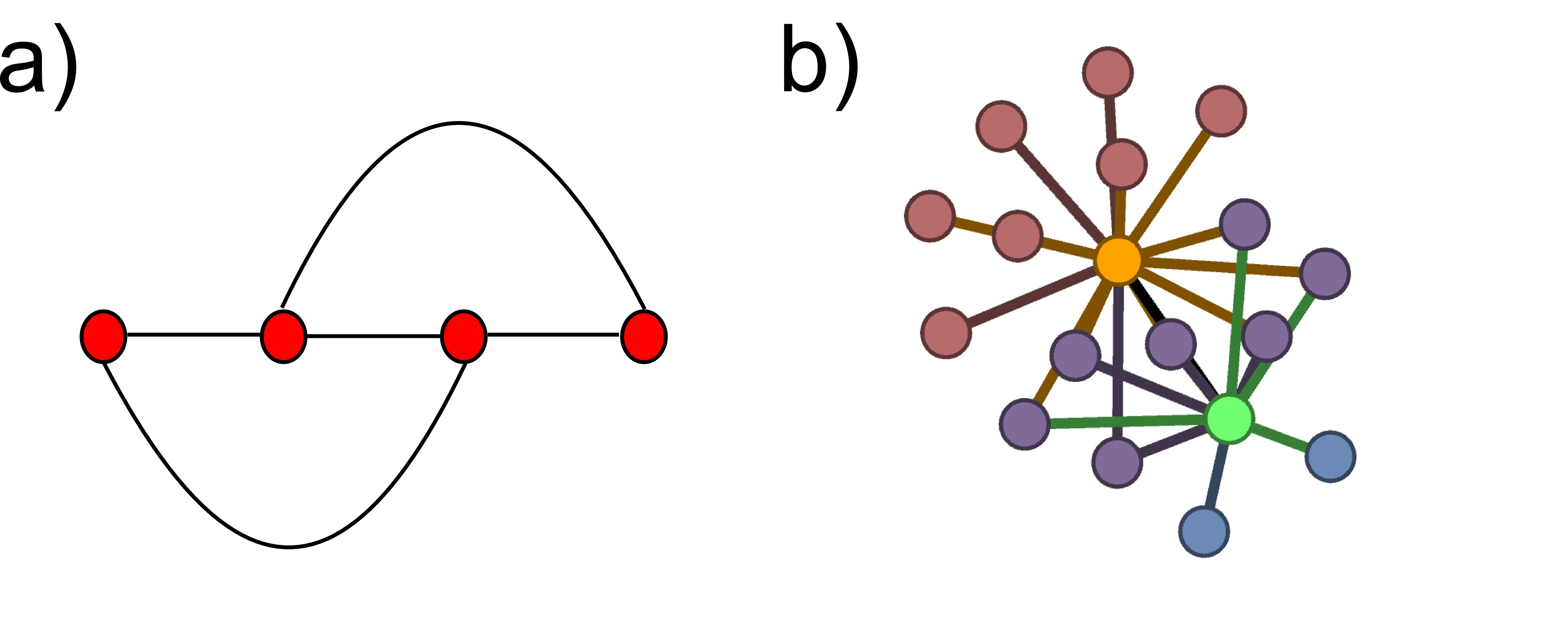}\protect\caption{Example of network motifs. a) An ``entangled cycle'' motif of size
four. b) A sample of the ``double-star'' motif, as found in the
real-world inter-AS topology. The star centers are in pink and green.
The nodes that are common to both stars are in purple. }
\end{figure}

In section \ref{sec:Dynamic-Analysis} we predicted the ubiquity of
two network motifs, the ``double star'' motif (Fig. \ref{fig:generalized star 2})
and the ``entangled cycles'' motif (Fig. \ref{fig:The-entangled-loops}).
We define the occurrence of a ``double star'' motif as the existence
of a a connected pair of nodes, each with degree greater than $m$,
designated as the centers, such that at least $m$ neighbors of one
center are also neighbors of the other center. We generated random
networks according to the Configuration Model (CM), in which each
node is given a number of stubs according to its degree, and stubs
are connected uniformly. Then, we evaluated the mean number of occurrences
of this motif in a random CM network with the same number of nodes
and the same degree distribution as the real inter-AS topology. For
$m=2$, we have found $28.8K$ occurrences in the real-world inter-AS
topology, whereas the mean number of occurrences in the random CM
network was only \textbf{$5.8K\pm1.3K$ }instances (the $\pm$ indicates
standard deviation). Namely, there are more than four times displays
of this subgraph in the Internet than in a random network with the
same degree distribution. Chebyshev's inequality provides a bound
on the \emph{p-value, $p<0.003$.} This low value indicates that it
is highly unlikely that a random CM network explains the frequent
appearance of this network motif. We have tested the prevalence of
this motif with other values of $m$ and the number of occurrences
is consistently a few times more than expected in a random CM network.
Our analysis suggests that reliability considerations is one of the
factors leading to the increased number of incidents.

The unexpected prevalence of the ``feedback loop'', which is a special
case of the ``entangled cycles'' motif, was first reported in \cite{Milo2002a}.
The ``feedback loop'' motif coincides with the \textquotedbl{}entangled
cycles\textquotedbl{} motif of length three, and in order to further
assess our results we tested for the occurrence frequency of the ``entangled
cycles'' motif of length four. We compared the number of occurrences
of this motif in the real-world Internet graph to the expected number
of occurrences in a random Configuration Model network. While the
number of instances of this motif in the Internet graph was $27.7M$,
the expected number of occurrences in the random network was only
$1.3M\pm0.8M$. As before, the abundance of this network motif, an
order of magnitude greater than expected (\emph{$p<0.001$}, a relaxed
bound based on Chebyshev's inequality) provides a positive indication
to the implications of survivability requirements.

In summary, we have provided both static and dynamic empirical evidence
that conform with our predictions, suggesting the importance of reliability
considerations on the structure and dynamics of the inter-AS topology.

\section{Conclusions }

While many studies have tried to model the Internet structure, the
list of works the explicitly address reliability considerations is
much shorter. Furthermore, most of the theoretical models of the Internet
structure and dynamics assume homogenous agents, while the Internet
is inherently heterogeneous, composed of a wide variety of entities
with different business models. In this work, we found that constructing
a model that includes these two factors, namely reliability and heterogeneity,
may provide important insights on the Internet structure and dynamics. 

We first rigorously formulated a model of a network formation game
in this context. Our model is flexible, and may be used in a wide
variety of settings. It allows for many variations and schemes, for
example situations in which failures are frequent or rare, or to account
for varying centrality of different types of players. The homogeneity
of the model also allows us to describe scenarios in which a fallback
route is required only to a subset of players. Indeed, a reasonable
AS policy is to require a backup routing path only to the Internet
backbone, rather than establishing fault-tolerant routing pathways
to every particular Autonomous System.

We established the \emph{Price of Reliability, }which measures the
excess social cost that is required in order to maintain network survivability
in an optimal stable equilibrium. Surprisingly, we showed that it
can be smaller than one, that is, that the additional survivability
constraints \emph{add }to the social utility. We have also showed
that reliability requirements have disparate effects on different
parts of the network. While it may support dilution in dense areas,
it facilitates edges formation in sparse areas, and in particular
it supports the formation of edges connecting minor players and major
players. 

In our dynamic analysis we have found the repetitive appearance of
small sub-graphs, or network motifs, namely the ``entangled cycles''
motif and the ``double star'' motif. Indeed, the number of appearance
of these motifs in the real Inter-AS topology surpassed the expected
number by a few folds, indicating that additional factors support
their formation, and as our analysis shows, survivability is one of
them. We have also predicted that the length of the minimal cycle
connecting a major player to a minor player should decrease in time.
This prediction, too, was verified by a dynamic data analysis.

Finally, while our analysis focuses on the inter-AS topology, it may
be applied to other networks as well, that are composed of heterogeneous,
rational agents that are required to maintain some reliability aspects.
Primary examples are trade networks and MVNO operators in the cellular
market.

In this work we have shown that a game theoretic analysis of network
formation, which encompasses heterogeneous agents and explicitly addresses
survivability concerns, holds promising results. Nevertheless, there
are many open questions, such as the following. How does the emerging
network handle more than a single failure? Which incentive mechanism
will promote increased reliability of the future Internet? What can
a comparative analysis of experimental results on different networks
tell us about the players' strategies in each network? These, and
many more, indicate that there is yet a lot to uncover in this intersection
between network formation, heterogeneity and reliability. 


\end{document}